\documentclass[aps,twocolumn,prl,superscriptaddress,showpacs]{revtex4}
\usepackage{epsfig,amsmath,amsbsy,amssymb}

\begin{document}

\title{Competitive localization of vortex lines and interacting bosons}
\author{J. Kierfeld}
\affiliation{Max-Planck-Institut f{\"ur} Kolloid- und 
  Grenzfl{\"a}chenforschung, 14424 Potsdam, Germany}
\author{V.M. Vinokur}
\affiliation{Argonne National Laboratory, 
   Materials Science Division, 
  9700 South Cass Avenue, Argonne, Illinois 60439, USA}
\date{\today}

\begin{abstract}
We present a theory for  the localization of three-dimensional vortex lines
or two-dimensional bosons with short-ranged repulsive interaction which are
competing for  a {\em single} columnar defect or potential well. For two
vortices we use a necklace model approach to find  a new kind of
delocalization transition between two different states with a single bound
particle. This {\em exchange-delocalization} transition is characterized by
the onset of vortex exchange on the defect for sufficiently weak
vortex-vortex repulsion or sufficiently weak binding energy corresponding to
high temperature. We calculate the transition point and order of the
exchange-delocalization transition.  A generalization of this transition to
arbitrary vortex number is proposed.
\end{abstract}

\maketitle

\paragraph{Introduction.}

Melting of the Bose glass, the low-temperature vortex phase in type-II
superconductors with columnar defects, remains a subject of constant
interest \cite{B03,kwok}. The interest is motivated not only by the appeal
and importance of understanding a basic phase transition of the vortex
system, but,  since the 3D vortex array is equivalent to a quantum 2D Bose
system \cite{NV}, vortex melting also offers a unique experimentally
accessible model to explore the interplay between disorder and interactions
in the delocalization transition of the corresponding strongly correlated 2D
quantum system.

Numerous experimental observations on Bose glass melting (see,  for example,
Refs.~\cite{Lia,kwok}) revealed a characteristic kink in the low-field
segment of the melting line suggesting a change of the melting mechanism. A
recent experimental study of BSCCO samples with a {\em very low} density
of columnar pins \cite{B03} allowed one to identify the low-field part of the
melting curve as  depinning transition from a single columnar defect driven
by vortex-vortex interactions. A theoretical study of the interacting boson
system with a low density of strong defects \cite{LV03} demonstrated the
possibility of an intermediate superfluid state where condensate and
localized bosons coexist. Furthermore, it was found in Ref.~\cite{LV03} that
interactions suppress localization and the increase of the boson density
results in a sharp {\em delocalization crossover} into a state where all
bosons are delocalized.

The model that is conventionally used in studies of quantum localization
can be viewed as an ensemble of (interacting) particles immersed in a random
field that can, in principle, localize or accomodate all particles; in
other words, there is a dense array of pinning centers struggling to
localize dilute, but interacting, particles. Reference \cite{LV03} proposed to
take an alternative approach and consider quantum particles or vortex lines
of high density competing for dilute traps or pinning sites. In this Letter,
we extend this approach and explore the regime of low particle densities,
i.e., a {\em finite} number of interacting quantum particles. 
A related quantum system that has been studied previously are two electrons
interacting in the region of a short-range attractive
 potential \cite{EAIL00}.

\begin{figure}[b]
\begin{center}
\epsfig{file=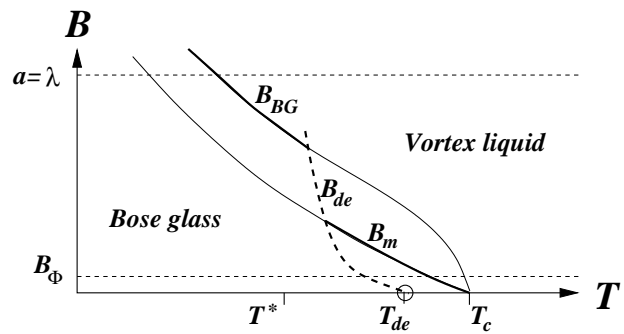,width=0.45\textwidth}
\end{center}
\caption{Schematic phase diagram in the $B$-$T$ plane. $B_m$ is the pristine
melting line, $B_{BG}$ the Bose glass melting line and $B_{de}$ the
exchange-delocalization crossover line which terminates at $B=0$ in a
genuine phase transition at temperature $T_{de}$ (circle). }
\label{phase}
\end{figure}

In this Letter, we investigate the formation of bound states in an ensemble
of $N$ strongly repulsive particles competing for a single attractive
potential well. In the related vortex system, this corresponds to
$N$ vortices competing for a single columnar defect, i.e., to a physical
situation where vortices outnumber columnar defects (magnetic fields
$B=NB_{\Phi}$
well exceed the matching field $B_{\Phi }=\Phi _{0}/a_{D}^{2}$ where $a_{D}$
is the average defect spacing and $\Phi _{0}$ the flux quantum). We focus on
the situation where mutual repulsion is strong enough to suppress
double-occupancy of the potential well and consider transversal
dimensionalities $d\leq 2$, for which a bound state for a single particle in
a symmetric potential well always exists (in the following, we use the
language of either particles or vortex lines at our will).
The main finding of this Letter is a new kind of  delocalization phase
transition driven by the {\em exchange} of the single 
bound particle with the $N-1$ unbound ones.
We first derive this result for $N=2$  and then propose 
a generalization for arbitrary $N$. Finally we obtain the
transition temperature $T_{de}$ for the exchange-delocalization transition
in the vortex system and discuss the resulting phase diagram 
Fig.~\ref{phase}. 
Contrary to Ref.~\cite{LV03} where the boson (vortex) density was finite,
we deal in this work with the genuine thermodynamic limit of {\em infinite}
system size but {\em finite} particle (vortex) number $N$ corresponding 
to $B,B_{\Phi }\approx 0$ with $N=B/B_{\Phi }$ finite. Thus the
exchange-delocalization phase transition emerges in the 
limit $B\to  0$ and replaces the crossover that was found 
in Ref.~\cite{LV03} for  macroscopic vortex density $B>0$.

\paragraph{The model.}

We describe a single vortex line in a sample of thickness $L$ interacting
with an attractive columnar defect by the Hamiltonian 
\begin{equation}
\mathcal{H}_{1}[\mathbf{r}(z)]=\int_{0}^{L}dz\left\{ \frac{1}{2}\varepsilon
_{l}(\partial _{z}\mathbf{r})^{2}+V_{D}(\mathbf{r}(z))\right\},
  \label{H1}
\end{equation}%
where $(z,\mathbf{r}(z))$ is the vortex trajectory in $1+d$ dimensions, the
magnetic field aligned with the $z$ axis, and $\varepsilon _{l}$ is the
stiffness or tilt modulus of a single vortex line; in an anisotropic
superconductor $\varepsilon _{l}\approx \varepsilon ^{2}\varepsilon _{0}\ln
\kappa $ where $\varepsilon _{0}=(\Phi _{0}/4\pi \lambda )^{2}$ is the
characteristic vortex line energy, $\lambda $ is the magnetic penetration
depth, $\kappa =\lambda /\xi \gg 1$, $\xi $ is the coherence length, 
and $\varepsilon $ is the anisotropy parameter. $V_{D}(r)$ is the pinning
potential from a single columnar defect positioned at $\mathbf{r}=0$. 
$V_{D}(r)$ falls off exponentially for $r>\lambda $ \cite{NV00} such that the
large scale behaviour of pinned vortex lines is well described using a
cylindrical pinning potential well 
\begin{equation}
V_{D}(\mathbf{r})=U_{D}~~\mbox{for}~~r<b_{D},~~~V_{D}(\mathbf{r})=0~~%
\mbox{for}~~r>b_{D}  
\label{Vd}
\end{equation}
with a potential depth $U_{D}\equiv -\varepsilon _{0}r_{D}^{2}/4\xi ^{2}$
and an effective radius $b_{D}\equiv \sqrt{2\xi ^{2}+r_{D}^{2}}$ where 
$r_{D}$ is the radius of the columnar defect \cite{NV}.

In $d\leq 2$ dimensions the line is {\em always} bound to the defect as can
be seen via mapping onto the ground-state problem of a quantum particle in a
potential $V_{D}(\mathbf{r})$
for large $L$ \cite{suris64}. 
Choosing the energy scale such that the
unpinned vortex line has a free energy $E_{0}=0$, the free energy per length 
$E_{1}<0$ of the bound vortex line is obtained as the ground state 
energy $E_{1}$ of the Schr{\"{o}}dinger equation 
\begin{eqnarray}
&&\left[ (T^{2}/2\varepsilon _{l})\nabla _{\mathbf{r}}^{2}-V_{D}(\mathbf{r})%
\right] \psi (\mathbf{r})=-E_{1}\psi (\mathbf{r}),  
\label{Schroeder} \\
&&E_{1}=U_{D}f(T/T^{\ast })
     \mbox{,~where}~~
 T^{\ast }\equiv b_{D}\sqrt{\varepsilon _{l}|U_{D}|}  
\label{E1}
\end{eqnarray}
is a characteristic crossover energy, and $f(x)$ a scaling function. For $d=2
$ and the pinning potential (\ref{Vd}), it has the asymptotic 
behavior $f(x)\approx 1-\mathcal{O}(x^{2})$ for $x\ll 1$ 
and $f(x)\approx x^{2}\exp (-2x^{2})/2$ for $x\gg 1$ \cite{NV}.

For $N>1$ we add the repulsive vortex interactions \cite{rev,NV} 
\begin{equation}
\mathcal{H}_N = \sum_{i=1}^N \mathcal{H}_1[\mathbf{r}_i(z)] 
 + \! \sum_{i\neq j =1}^{N}\int_0^L dz 
        V(|\mathbf{r}_i(z)\! -\!\mathbf{r}_j(z)|),
\label{Hint}
\end{equation}
where $V(r) = 2\varepsilon_0 K_0(r/\lambda)$, and $K_0$ is the Bessel
function. Double-occupancy of the defect over an extended length is
energetically disfavored if $\varepsilon_0 \gg |E_1|$, which is always the
case at high enough temperatures. We also focus on the regime of large
vortex spacing $a \gg \lambda$.

\begin{figure}[b]
\begin{center}
\epsfig{file=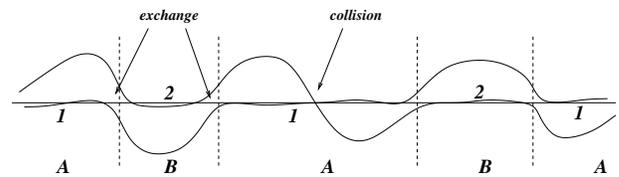,width=0.45\textwidth}
\end{center}
\caption{Two particles binding alternately to a single columnar defect. The
particle binding to the defect can be exchanged in localized events and
there are rare collisions between the free and the bound particle. }
\label{necklace}
\end{figure}

\paragraph{Exchange-delocalization transition for N=2.}

We investigate the localization behavior for the case $N=2$ making use of a 
{\em necklace model} approach \cite{F84}. As the double-occupancy of the
defect is energetically unfavorable, the accessible configurations of the
vortex lines consist of a necklacelike succession of two possible
configurations \textit{A} and \textit{B}, see Fig.~\ref{necklace}. In the
configuration \textit{A}, the line 1 is bound to the defect with the binding
free energy $E_{1}<0$ and the transversal 
localization length $\xi _{\perp}=T/\sqrt{|E_{1}|\varepsilon _{l}}$, 
whereas the line 2 is essentially in
the unbound state with the free energy $E_{0}=0$ and experiences rare
collisions with the bound line 1. As we assume $\lambda \ll a$, the unbound
line is exploring the region $r>\lambda $ of  exponentially weak
repulsion, whereas collisions occur at $r<\lambda $. The configuration 
\textit{A} ends in an exchange event where the endpoints of the unbound line
2 attach to the defect again; see Fig.~\ref{necklace}. At these exchange
points the configuration \textit{A} can connect to the 
configuration \textit{B} where the roles of the particles are exchanged.

First, we estimate the energy cost of the localized collision and exchange
events. In the presence of the repulsion (\ref{Hint}), each return of line 2
to the bound line 1 will cost an additional collision repulsion energy $E_r$
which is determined by optimizing the sum of elastic and repulsive energies
for a contact of  length $\ell_r$ over which the typical line spacing is of
the order $|\Delta \mathbf{r}| \simeq \lambda$, $E_r \simeq \varepsilon_l
\lambda^2/\ell_r + \ell_r \varepsilon_0$: 
\begin{equation}
\ell_r \simeq \lambda \sqrt{\varepsilon_l/\varepsilon_0},~~~~
 E_r \simeq \lambda \sqrt{\varepsilon_l\varepsilon_0},  
\label{Er}
\end{equation}
and $v_r\equiv \exp({-E_{r}/T})<1$ defines the Boltzmann-factor associated
with each collision. Similarly, we estimate the energy cost $E_{ex}$ of a 
localized exchange by optimizing  the sum of the elastic energy and the loss
of binding energy  $|E_1|\ell_{ex} $  for a contact of 
  length $\ell_{ex}$ over which the typical 
line spacing is of the order $|\Delta \mathbf{r}|
\simeq \lambda$, $E_{ex} \simeq \varepsilon_l \lambda^2/\ell_{ex} + \ell_c
|E_1|$. This  gives 
\begin{equation}
\ell_{ex} \simeq \lambda \sqrt{\varepsilon_l/|E_1|},~~~~ 
E_{ex} \simeq \lambda \sqrt{\varepsilon_l|E_1|}, 
\label{Eex}
\end{equation}
and $v_{ex}\equiv \exp({-E_{ex}/T})<1$ is the Boltzmann-factor associated
with each localized exchange.

Now we address the statistical mechanics problem of summing over all vortex
line configurations. Adopting a coarse-grained description focusing on
scales $r\gtrsim \lambda $ for transversal vortex fluctuations, we
discretize the vortex system into segments of length $l=\lambda
^{2}\varepsilon _{l}/T$ in the $z$-direction. In what follows, we calculate
the grand-canonical partition sum $ G(z)=\sum_{n}Z(n)z^{n}$ where $Z(n)$ is
the partition sum for a system of length $L=nl$ and $z$ is the 
fugacity. $\tilde{G}(E)=G(\exp {(lE/T)})$ 
is the Green's function at energy $E$ for the
corresponding two-particle quantum problem. The free energy density $f$ of
the system is determined by the real singularity $z_{0}$ of $G(z)$ closest
to the origin by the relation $z_{0}=\exp (lf/T)$. If $G_{A}(z)$ 
and $G_{B}(z)$ are the partition sums for configurations A and B,
respectively, the full partition sum  is obtained by summing over
all alternating configurations $G_{A}$, $G_{B}$, $G_{A}v_{ex}G_{B}$, 
$G_{B}v_{ex}G_{A}$, $G_{A}v_{ex}G_{B}v_{ex}G_{A}$,\ldots, separated by
particle exchanges with Boltzmann-factor $v_{ex}$. Summing up the resulting
geometric series we obtain 
\begin{equation}
\!\!G(z)=\left. \frac{G_{A}+G_{B}+2v_{ex}G_{A}G_{B}}{1-v_{ex}^{2}G_{A}G_{B}}%
\right\vert _{z}=\frac{2G_{A}(z)}{1-v_{ex}G_{A}(z)},  
\label{Gz}
\end{equation}%
where we used $G_{A}=G_{B}$ because both configurations are related by a
mere particle exchange. [Boundary effects are irrelevant; we allow either 
\textit{A} or \textit{B} at the ends of the defect in (\ref{Gz})]. According
to Eq.\ (\ref{Gz}), the singularity determining the free energy of the system
is given either by the singularity of $G_{A}(z)$ corresponding to the state
where the \emph{same} line is always bound or by the solution 
of $1=v_{ex}G_{A}(z)$ corresponding to alternating bound particles. 
The {\em exchange-delocalization transition} between these 
two states occurs if both
singularities occur at the same value of $z$.

To move further, we calculate the grand-canonical partition 
sum $G_{A}(z)$. In the absence of the interline repulsion, the canonical
partition sum in 
configuration \textit{A}, $Z_{A}(n)=Z_{1}(n)Z_{2}(n)$, is a product
of the partition sum of the bound line 1, $Z_{1}(n)=\exp
(-nlE_{1}/T)$ [see Eq.\ (\ref{E1})], and of the free line 2, which is attached
with its end-points to the defect. This restriction leads 
to $Z_{2}(n)=p_{n}\exp (-nlE_{0}/T)$ where $p_{n}$ is the probability 
for the return of the unbound line to the 
defect and $E_{0}=0$ is its free energy
per unit length. In the absence of the repulsion this return probability is
given by the return probability of a random walk, $p_{n}\simeq n^{-d/2}$,
for large $n$. Then the partition function $G_{A}(z)$ is related to the
generating function $P(z)=\sum_{n}p_{n}z^{n}$ for these return probabilities
by $G_{A}(z)=P(wz)$ where $w\equiv \exp (-lE_{1}/T)=\exp (-\lambda
\varepsilon _{l}E_{1}/T^{2})$. For $1-z\ll 1$, $P(z)\sim
(1-z)^{d/2-1}$ 
for $d<2$ and $P(z)\sim -\ln (1-z)$ for $d=2$. Including the 
Boltzmann-factor $v_{r}$ in the random walk of 
the unbound line for each collision with the
repulsive bound line localized at the defect [see Eq.~(\ref{Er})]
leads to a modified generating function $P_{r}(z)$: 
\begin{equation}
P_{r}(z)=P(z)+P(z)v_{r}P_{r}(z)-P(z)P_{r}(z).  
\label{Pr}
\end{equation}
In this relation the contributions from the repulsion-free walks with
Boltzmann-factor $1$ are subtracted and the corresponding
term with modified Boltzmann-factor $v_{r}$ is added  on
the right hand side recursively for each collision. With this modification
due to collisions, we finally obtain 
\begin{equation}
G_{A}(z)\!=\!P_{r}(wz)\!=\!\frac{P(wz)}{1\!+\!(1-v_{r})P(wz)},
\,w\equiv e^{-lE_{1}/T}  
\label{GA}
\end{equation}
for the grand-canonical partition function. $G_{A}(z)$ has a 
singularity at $z=1/w$ corresponding to a free energy per 
length $f=E_{1}$ identical to that of a single bound particle 
because the second unbound particle has $E_{0}=0$. The 
function $P(wz)$ diverges upon approaching $z=1/w$ and, thus,
we find $G_{A}(1/w)=P_{r}(1)=(1-v_{r})^{-1}$ at the singularity.

Now we turn to the exchange-delocalization transition determined by the
singularities of (\ref{Gz}). We have already found that the 
singularity of $G_A(z)$ (at $z=1/w$ corresponding to the 
free energy $f = E_1$) describes,
indeed, a single bound particle, i.e., a state with always the {\em same}
line bound. We have argued above that there can be a real singularity closer
to the real axis which is given by the solution of the 
equation $1=v_{ex}G_A(z)$ and which corresponds to the 
phase with exchanging bound particle. From the functional 
form (\ref{GA}) of $G_A(z)$, one readily
verifies that this singularity is indeed the one  closer to the origin and
therefore representing the thermodynamically stable phase provided 
\begin{equation}
v_{ex} \ge v_{ex,c} = G_A^{-1}(1/w) = 1-v_r ~.  
\label{transition}
\end{equation}
According to (\ref{transition}), the exchange-delocalization phase transition
occurs at the critical temperature $T_{de}$ that is obtained 
from $\exp(-E_{r}/T_{de}) \approx E_{ex}/T_{de}$ where we 
assumed that $E_{ex} \ll T_{de}$ because $T_{de} > T^*$. Using 
estimates (\ref{Er}) and (\ref{Eex}),
the asymptotics of the function $f(x)$ for $d=2$, 
and $\lambda/b_D= \kappa/\sqrt{2}$, we arrive at 
\begin{equation}
T_{de} \approx T^* \kappa^{1/3} 
 \label{Td}
\end{equation}
for the delocalization transition temperature $T_{de}$ in the vortex system.
The transition takes place in the regime $T>T^*$  where a single vortex is
only weakly bound to the defect \cite{NV}; see Fig.~\ref{phase}. Note that
both Eqs.\ (\ref{E1}) and (\ref{Td}) are self-consistent equations for $T^*$
and $T_{de}$, respectively, due to the temperature dependence 
of  $\xi$ and $\lambda$. Furthermore, it can be 
shown by expanding the  equation $1=v_{ex}G_A(z)$ about 
the transition point, that the 
exchange-delocalization  transition is continuous for all $d\le 2$ and of 
infinite order for $d=2$.

We expect our results to apply to all short-ranged potential wells and
particle interactions that decay faster than $1/r^2$ for large 
separations $r$ \cite{L89}. More 
realistic pinning potentials
contain an intermediate $1/r^2$-behaviour on scales $\xi \ll r \ll \lambda$ 
\cite{rev,LV03}, which slightly changes the function $f(x)$ and thus
the exact value of $T_{de}$ but not the universal properties of the
delocalization transition.

\paragraph{General N.}

We start from the exchange-delocalized state where all $N$ vortices share
the defect and consider the instability with respect to the exclusion of one
of the vortices from the exchange. To this end we introduce $N$ states
(analogously to the states A and B for $N=2$), where one vortex is unbound,
i.e., excluded from the exchange, whereas the other $N-1$ vortices share the
defect. Then the necklace is a succession of possible states $i=1,...,N$
each of which has a grand-canonical partition 
sum $G_{i}(z)=P(w_{N}z)/(1+[1-v_{r}(N)]P(w_{N}z))$ 
where $v_{r}(N)\equiv \exp (-\sqrt{N-1}E_{r}/T)$ is the 
Boltzmann-factor due to the enhanced repulsion
from the $N-1$ vortices sharing the columnar defect. Similarly, $w_{N}\equiv
\exp (-lf_{N-1}/T)$ is the Boltzmann-factor for the binding free energy of
the $N-1$ vortices sharing the defect, which we approximate 
by $f_{N-1}\approx E_{1}$ or $w_{N}\approx w=\exp (-lE_{1}/T)$. 
Considering the generalized exchange between these $N$ states 
and noting that exchange of
the single bound particle is  associated with the 
Boltzmann-factor $v_{ex}$, we arrive at the generalization of (\ref{Gz}): 
\begin{equation*}
G(z)=\frac{NG_{i}(z)}{1-(N-1)v_{ex}G_{i}(z)}~.
\end{equation*}
The transition point for the exclusion of one particle from the exchange is
given by the relation $v_{ex,c}(N)=[1-v_{r}(N)]/(N-1)$. As $v_{ex,c}(N)$
decreases for increasing $N$, also states with $N-1$ or less exchanging
particles become unstable for $v_{ex}<v_{ex,c}(N)$. This leads to the
conclusion that particle exchange entirely stops at this point, and we thus
identify $v_{ex}=v_{ex,c}(N)$ as the exchange-delocalization transition
point of the $N$-particle system. For $N=2$ our result 
reduces to Eqs.\ (\ref{transition}) and (\ref{Td}), whereas we find 
\begin{equation}
T_{de}(N)\approx \left\{ 
\begin{array}{ll}
T^{\ast }\ln ^{1/2}\left( \kappa /\ln N\right)  & \mbox{for}~~\ln N<\kappa 
\\ 
T^{\ast }\kappa /\ln N & \mbox{for}~~\ln N>\kappa 
\end{array}%
\right.   \label{TdN}
\end{equation}%
for large $N=B/B_{\Phi }$. Note that our approach is limited to the regime
of large vortex spacing $a\gg \lambda $ or $N\ll a_{D}^{2}/\lambda ^{2}$.
The order of the transition is the same as for $N=2$; i.e., the transition
is of infinite order for $d=2$.

\paragraph{Phase diagram.}

In real vortex systems our results hold for the limit $B_{\Phi }\approx 0$.
This implies also a vanishing vortex density $B\approx 0$ if $N=B/B_{\Phi }$
is fixed. A finite vortex density $\rho >0$ corresponds to a system of the
finite size $\propto 1/\rho ^{1/2}$ which has no genuine phase transition.
We thus conclude that the delocalization crossover line of Ref.~\cite{LV03}
terminates in the genuine exchange-delocalization \emph{transition point} at 
$B\approx 0$, which is given by Eqs.\ (\ref{Td}) or (\ref{TdN}). For
macroscopically large $N$ the approximation of localized, well-separated
exchange and repulsion events will break down and our low-density approach
will become invalid whereas the description by a condensate of bosonic
particles used in Ref.~\cite{LV03} works increasingly well in the
high-density regime. The exact form of the crossover between both
descriptions is an open question. In both descriptions the delocalization
line $B_{de}(T)$ drops exponentially with temperature [see (\ref{TdN})] such
that it intersects with both the pristine melting line $B_{m}(T)$ and the
Bose glass melting line $B_{BG}(T)$; see Fig.~\ref{phase}. Beyond the
delocalization line, vortex line exchange at the defects sets in,
 which leads to line wandering and a liquidlike behavior even in 
the presence of columnar defects, which become irrelevant. 
Therefore, the relevant melting
line is the pristine melting line for $T>T_{de}(B)$ in the delocalized
phase, whereas it is the Bose glass melting line in the localized phase.
Therefore, there exists a range of magnetic fields where the vortex lattice
melts by undergoing the delocalization transition \cite{LV03} as shown in
Fig.~\ref{phase}. The resulting phase diagram is in good agreement with the
experimental results regarding the melting of \textquotedblleft
porous\textquotedblright\ vortex matter \cite{B03}.

\paragraph{Conclusion.}

In conclusion, we have shown that the competitive localization of particles
with mutual short-range repulsion by a short-range attractive defect
leads to the existence of a genuine phase transition, the {\em 
exchange-delocalization transition}, which marks the onset of particle
exchange at the defect. We have investigated the exchange-delocalization
transition in a system of vortices with columnar defects or interacting
bosons with localizing defects at low density. We expect
exchange-localization transitions to play an important role in various other
systems where competitive localization occurs. In the introduction, we have
already mentioned the quantum mechanical system of two interacting electrons
competing for an attractive potential \cite{EAIL00} that could be realized
by a quantum dot. We also expect  competitive localization to be relevant
for biopolymers, for example it applies to the competitive binding of two
identical single strands of DNA to a single complementary DNA strand as it
is important for DNA microarray engineering.

\paragraph{Acknowledgments.}

This research is supported by the US DOE Office of Science under contract
No. W-31-109-ENG-38.


\end{document}